\documentclass[12pt]{article}

\usepackage{amsthm,amsmath,natbib}
\RequirePackage{hyperref}

\usepackage{amsfonts}
\usepackage{graphicx}
\usepackage{comment}
\usepackage{mathrsfs}

\newtheorem{theorem}{Theorem}[section]

\begin{document}

\def\spacingset#1{\renewcommand{\baselinestretch}%
{#1}\small\normalsize} \spacingset{1}

  \title{\bf Interleaved lattice-based maximin distance designs}
  \author{Xu He\hspace{.2cm}\\
    Academy of Mathematics and System Sciences, \\Chinese Academy of Sciences}
  \maketitle

\begin{abstract}
We propose a new method to construct maximin distance designs with arbitrary number of dimensions and points. 
The proposed designs hold interleaved-layer structures and are by far the best maximin distance designs in four or more dimensions. 
Applicable to distance measures with equal or unequal weights, 
our method is useful for emulating computer experiments when a relatively accurate priori guess on the variable importance is available. 
\end{abstract}


\spacingset{1.45} % DON'T change the spacing!

\section{Introduction}
\label{sec:intro}

Computer experiments have become powerful tools to simulate real systems for which actual experimentation is expensive. 
Space-filling designs whose points are in some sense ``uniformly'' scattered in the design space are common choices for computer experiments~\citep{Sacks:1989,Santner:2003}. 
Two notable criteria for uniformity are separation and fill distances.   
The $L_2$ separation distance of a design $D$ is the minimal $L_2$ distance among pairs of design points, 
\begin{equation}\label{eqn:sep}
\rho(D) = \min_{x,y \in D} \|x-y\|_2, 
\end{equation} 
and the $L_2$ fill distance of a design $D$ in $[0,1]^p$ is the supremum of predictive distance of any position in the design space, 
\[ \sup_{y\in [0,1]^p} \left( \min_{x\in D} \|y-x\|_2 \right). \]
Designs with maximum separation distance and minimum fill distance are called maximin and minimax distance designs, respectively~\citep{Johnson:1990}. 
\citet{Haaland:2016} and~\citet{Wang:2017} studied on the broad principles for experimental design of computer experiments. 
They concluded that space-filling designs with high separation distance and low fill distance are appealing because they ensure accurate Gaussian process emulation of computer experiments. 

Since the two distance-based criteria are largely non-conflicting, in practice researchers usually use one of them to construct space-filling designs. 
The separation distance criterion is used much more often than the fill distance criterion, presumably because of the following reasons: 
Firstly, maximin distance designs are asymptotically D-optimal for Gaussian process models~\citep{Johnson:1990}. 
Secondly, the separation distance controls the numerical error in Gaussian process emulation~\citep{Haaland:2016}, which is crucial for large sample experiments. 
Thirdly, Gaussian process models fitted from maximin distance designs are robust to simulation errors~\citep{Siem:2007}. 
Fourthly, it is computationally and theoretically easier to obtain the separation distance of designs than the fill distance. 
In this paper we focus on constructing designs in $[0,1]^p$ with high separation distance. 

The problem of constructing maximin distance designs has been studied by many researchers. 
Virtually all existing methods treat it as a numerical optimization problem for which the objective function is the separation distance. 
The website \url{http://www.packomania.com/} lists the best-known maximin distance designs for $p=2$ and $3$, which are the best from 32 algorithms. 
These designs are very good already and there is little room for improvement. 
On the other hand, the numerical optimization problem becomes much more difficult for higher $p$. 
Although algorithms to construct maximin distance designs in general $p$ are available~\citep{Trosset:1999,Stinstra:2003,Mu:2017}, 
designs generated from them are far from optimal unless the sample size is very small. 

In this work, we propose to generate maximin distance designs from interleaved lattices. 
Interleaved lattices are sets of points with special layered structure and were shown useful in constructing minimax distance designs~\citep{ILmMD}. 
As we shall discuss in Section~\ref{sec:lattice}, designs with high separation distance tend to hold interleaved-layer structures. 
Figure~1 depicts an interleaved lattice-based design with $p=3$ and sample size $n=148$. 
Its separation distance is 0$\cdot$2430, better than that of the design listed in \url{http://www.packomania.com/} by 0$\cdot$002. 
Lattice or layered structures have been explored in other works to construct designs with other distance-based uniformity criteria~\citep{Zhou:2015,ILmMD,RSPD,Guiban:2017,Zhou:2017,Xiao:2017,Xiao:2017+}. 

We propose three algorithms to construct interleaved lattice-based maximin distance designs. 
Our algorithms try various interleaved lattices and scale parameters to find a design with high separation distance.  
By focusing on only interleaved lattice-based designs the search space is greatly reduced. 
In addition, we exploit mathematical properties of lattices to further simplify the search. 
Consequently, we can efficiently construct designs with excellent separation distance for general $p$ and $n$. 
Numerical results suggest our proposed designs usually have at least 0$\cdot$1 higher separation distance than those generated from numerical optimization algorithms for general $p$. 

\begin{figure} \label{fig:exp} 
\begin{center}
\includegraphics[width=13cm]{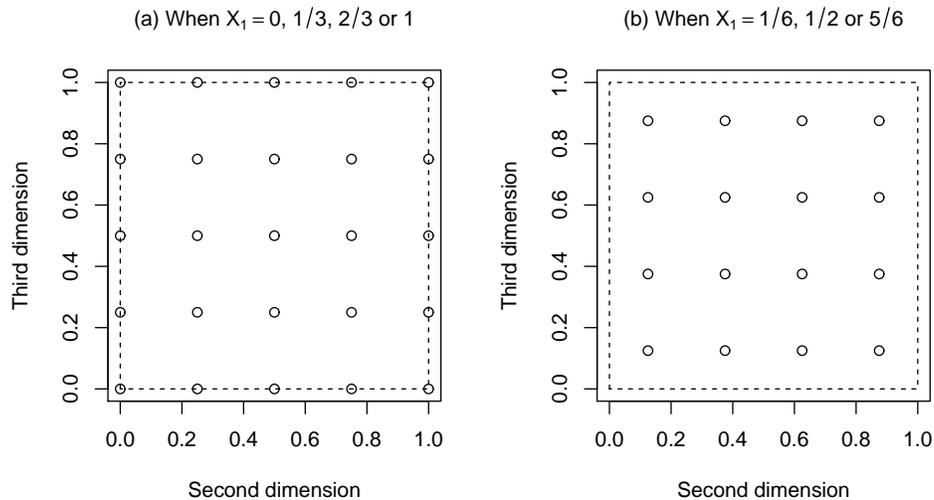}
\caption{The second and third dimensions of the interleaved lattice-based maximin distance design with $p=3$ and $n=148$ when the value of the first dimension $X_1$ is (a) 0, $1/3$, $2/3$, 1 and (b) $1/6$, $1/2$, $5/6$. This design has seven distinct values for the first dimension and its second and third dimensional values are the same when $X_1 = 0, 1/3, 2/3$, and 1 and when $X_1 = 1/6, 1/2$, and $5/6$, respectively.}
\end{center}
\end{figure}

When some variables have much stronger impact on the response than others and the relative importance of variables is known, 
it is advantageous to use 
a weighted $L_2$ distance measure~\citep{Wang:2016}, 
\begin{equation}\label{eqn:distance}
d(x,y,w) = d(y-x) = \left\{ \sum_{k=1}^p \left( w_k|x_k-y_k| \right) ^2 \right\}^{1/2}, 
\end{equation}
where $x_k$ and $y_k$ denote the $k$th dimensional value of $x$ and $y$, respectively, 
and $w_k$ quantifies the importance of the $k$th variable. 
Variables having stronger impact on the response should be assigned with higher weight. 
Throughout this paper, we assume $w=(w_1,\ldots,w_p)$ is known and use the distance measure in (\ref{eqn:distance}) to construct designs. 
We remark that generating a maximin distance design in $[0,1]^p$ with the weighted distance measure in (\ref{eqn:distance}) is equivalent to generating a maximin distance design in $\prod_{k=1}^p [0,w_k]$ with the unweighted distance measure and transforming the design to the $[0,1]^p$ space. 
Numerical results provided in Section~\ref{sec:simu} suggest our proposed designs are more suitable than maximin distance Latin hypercube designs~\citep{Morris:1995} in emulating computer experiments when 
relatively accurate prior knowledge on variable importance is available.

\section{Theoretical results}
\label{sec:lattice}

In this section we give useful theoretical results on interleaved lattice-based designs, focusing on their separation distance properties. 
We begin by reviewing the definition of interleaved lattices. 
A set of points $L \subset \mathbb{R}^p$ is called a lattice in $p$ dimensions with full rank and $G$ a generator matrix of $L$ if $G$ is a nonsingular $p\times p$ matrix and 
$ L = \left\{ a G : a \in \mathbb{Z}^p \right\}$. 
For instance, $\mathbb{Z}^p$ is called the $p$-dimensional integer lattice and $I_p$, the $p$-dimensional identity matrix, is one generator matrix of $\mathbb{Z}^p$. 
Another example is the even integer lattice $\mathbb{E}^p$, where $\mathbb{E}$ is the set of even integers. 
One generator matrix of $\mathbb{E}^p$ is $2I_p$.  
Suppose $L_1$ and $L_2$ are two lattices in $p$ dimensions with full rank. 
If $L_1$ is a subset of $L_2$, we call $L_1$ a sublattice of $L_2$ and $L_2$ a superlattice of $L_1$. 
Clearly, $\mathbb{E}^p$ is a sublattice of $\mathbb{Z}^p$. 
See \citet{Conway:1998} for a comprehensive review of lattices with distance properties. 

Layered lattices are lattices that can be partitioned into layers based on each dimension. 
Interleaved lattices are layered lattices with repeated or alternated layers. 
For instance, we can partition the design in Figure~\ref{fig:exp} into seven, nine, and nine alternated layers based on the first, second, and third dimensions, respectively. 
Formally, a lattice $L$ is called a standard interleaved lattice if 
$\mathbb{E}^p \subset L \subset \mathbb{Z}^p$ and $\{x: x\in L, x_k=1\} \neq \emptyset$ for any $k=1,\ldots,p$. 
For a $p$-vector $c=(c_1,\ldots,c_p)$, let $c \otimes L = \{(c_1x_1,\ldots,c_px_p): x \in L\}$ and $L \oplus c = \{ x+c : x\in L\}$. 
A set is called an interleaved lattice if it can be expressed as $\left( b \otimes L \right) \oplus c$ where $L$ is a standard interleaved lattice. 
Clearly, any lattice $L$ such that $\mathbb{E}^p \subset L \subset \mathbb{Z}^p$ is an interleaved lattice. 
Treating dimension permuted lattices as different lattices, we find 2, 6, 26 and 158 distinct standard interleaved lattices in 2, 3, 4 and 5 dimensions, respectively, after exhausting all possibilities. 
For higher $p$, there exist many more types of standard interleaved lattices. 

Two important quantities of lattices $L$ with $\mathbb{E}^p \subset L \subset \mathbb{Z}^p$ are their $q$ and $r$ values. 
Let $e_k$ denote the $p$-vector whose $k$th entry is one and other entries are zeros, 
$\vert T \vert$ denotes the cardinality of the set $T$, 
$q(L) = \log_2 \vert L \cap \{0,1\}^p \vert$ and 
$r(L) = \vert \{k:e_k \in L\} \vert$. 
The $q$ and $r$ are nonnegative integers satisfying either $r=q=p$ or $0\leq r<q<p$~\citep{ILmMD}.  

A set $D$ is called an $L$-based design in $[0,1]^p$ if $D$ can be expressed as 
\begin{equation}\label{eqn:LbasedD}
D = \left\{ \left( b \otimes L \right) \oplus c \right\} \cap [0,1]^p  
\end{equation}
with a pair of $p$-vectors $b$ and $c$, i.e., it consists of the rescaled and translated lattice points that lie in $[0,1]^p$ . 
See \citet{ILmMD} for more results on interleaved lattices. 

Next, we give some theoretical results on the size and separation distance of interleaved lattice-based designs. 
Let $s_k$ denote the number of distinct values of the $k$th dimension of $D$. 
Following~\citet{ILmMD}, call $s=(s_1,\ldots,s_p)$ the span vector of $D$ 
and write $s^{-1} = (s_1^{-1},\ldots,s_p^{-1})$.
Let $\lfloor z \rfloor$ and $\lceil z \rceil$ denote the largest integer no greater than $z$ and the smallest integer no lower than $z$, respectively. 
Clearly, 
\[ 
s_k = \lfloor (1-c_k)/b_k \rfloor - \lceil -c_k/b_k \rceil + 1. 
\]
Theorem~\ref{thm:m} below shows that the size of $D$, denoted as $m(D)$, depends strongly on $q(L)= \log_2 \vert L \cap \{0,1\}^p \vert$ and $s$. 

\begin{theorem} \label{thm:m}
Suppose $D$ is generated from a lattice $L$ via (\ref{eqn:LbasedD}) and $\mathbb{E}^p \subset L \subset \mathbb{Z}^p$. 
Then
\begin{equation}\label{eqn:upper:quick}
2^{q-p} \prod_{e_k \notin L} \left( 2 \lfloor s_k/2 \rfloor \right) \prod_{e_k\in L} s_k \leq m(D) \leq 2^{q-p} \prod_{e_k \notin L} \left( 2 \lceil s_k/2 \rceil \right) \prod_{e_k \in L} s_k . 
\end{equation}
Furthermore, 
among all $L$-based designs with the same $s$, the maximal $m(D)$ can be attained by a $D$ with $0_p \in D$. 
\end{theorem}

All proofs are provided in the appendix.
From Theorem~\ref{thm:m}, the size of $D$ is largely determined by $q$ and $s$. 
Loosely speaking, the average $m(D)$ is $2^{q-p} \prod_{k=1}^p s_k$. 
Although we can always compute $m(D)$ from $L$, $b$ and $c$, (\ref{eqn:upper:quick}) gives an upper bound of $m(D)$ that is very fast in computation. 
Let $1_p$ and $0_p$ denote the $p$-vector with ones and zeros, respectively, 
and  
\begin{equation}\label{eqn:L0}
L_0 = \{ x: x\in L, x_k=0  \mbox{ or } ( x_k=1 \mbox{ and } e_k \notin L ), k=1,\ldots,p \} . 
\end{equation}
Clearly, $L$ is determined by $L\cap \{e_1,\ldots,e_p\}$ and $L_0$. 
Theorem~\ref{thm:sep} below characterizes the $b$, $c$ and separation distance of interleaved lattice-based designs. 

\begin{theorem} \label{thm:sep}
For any given $p$ and $n\geq 2$, among $D$s generated via (\ref{eqn:LbasedD}) with $m(D)\geq n$, the highest $\rho(D)$ can be attained by an $(L,b,c)$ combination in (\ref{eqn:LbasedD}) with $c=0_p$ and $s_k-1=1/b_k \in \mathbb{N}$, $k=1,\ldots,p$, and thus 
\begin{equation}\label{eqn:D}
D = (s-1_p)^{-1} \otimes \left\{ L \cap \left( \prod_{k=1}^p \{0,\ldots,s_k-1\} \right) \right\}.
\end{equation}
Furthermore, for such $D$, 
\begin{equation}\label{eqn:rho}
\rho(D) = \min\left[ \min_{x\in L_0, x\neq 0_p} d\{(s-1_p)^{-1} \otimes x\}, \min_{e_k\in L} \{w_k/(s_k-1)\}, \min_{s_k> 2} \{2w_k/(s_k-1)\} \right]. 
\end{equation}
\end{theorem}

In light of Theorem~\ref{thm:sep}, in constructing maximin distance designs we only consider designs that can be expressed as in (\ref{eqn:D}). 
Such designs are determined by $L$ and $s$ and contain $0_p$, the origin. 
We use $D(L,s)$, $\rho(L,s)$ and $m(L,s)$ to denote the design generated by (\ref{eqn:D}), its separation distance and its size, respectively. 
It is faster to compute the separation distance via (\ref{eqn:rho}) than via (\ref{eqn:sep}). 
Besides, (\ref{eqn:rho}) provides insights on what $L$ leads to maximin distance designs, which we shall further discuss in Section~\ref{sec:construction2}. 

Clearly, designs generated via (\ref{eqn:D}) have many points on the boundary of $[0,1]^p$. 
We can as well construct interleaved lattice-based designs via
\begin{equation}\label{eqn:tildeD}
\tilde D(L,s) = s^{-1} \otimes \left[ \left\{ L \cap \left( \prod_{k=0}^p \{0,\ldots,s_k-1\} \right) \right\} \oplus (1_p/2) \right]. 
\end{equation}
Let the Voronoi cell of a point $x$ in a design $D$ be the region nearer to $x$ than other design points, given by 
 $\text{Vor}(x) = \left\{ y \in \mathbb{R}^p :  \| y-x \|_2 \leq \| y- \tilde x \|_2 \mbox{ for any } \tilde x \in D \right\}$.   
Points of $\tilde D(L,s)$ are roughly located at the center of their Voronoi cells, and the Voronoi cells have roughly equal volumes. 
Consequently, points of $\tilde D(L,s)$ represent $[0,1]^p$ better than that of $D(L,s)$ generated via (\ref{eqn:D}) in not exaggerating the near-boundary regions. 
Thus, they can be seen as support points and may be useful in some applications~\citep{Mak:2017}. 
On the other hand, $D(L,s)$ has higher separation distance and may be more suitable to the emulation problem for which denser points in the near-boundary regions is desired~\citep{Dette:2010}. 

Obviously, for the same $L$, $m(L,s)$ increases and $\rho(L,s)$ decreases as the elements of $s$ grow. 
For the same $s$, lattices with higher $q$ almost always lead to higher $m(L,s)$ and lower $\rho(L,s)$. 
Theorem~\ref{thm:rep} below shows that $m(L,s)$ is also related to $r = \vert \{k:e_k\in L\} \vert$. 

\begin{theorem} \label{thm:rep}
For any $s\in \mathbb{N}^p$ and $0\leq z_1 \leq z_2 < z_3 <p$, 
\[ \max_{q(L)=z_3,r(L)=z_2} m(L,s) \geq \max_{q(L)=z_3,r(L)=z_1} m(L,s), \]
where both maximums are over lattices $L$ with $\mathbb{E}^p\subset L \subset \mathbb{Z}^p$. 
\end{theorem}

From Theorem~\ref{thm:rep}, for the same $s$, lattices with the same $q$ but higher $r$ tend to lead to higher $m(L,s)$. 
From our experience, the impact of $r$ is much weaker than that of $s$ and $q$. 
Furthermore, for the same $s$, the $m(L,s)$ does not vary much among lattices with the same $q$ and $r$. 
Theorems~\ref{thm:m}-\ref{thm:rep} are useful for finding proper $L$ and $s$ in generating interleaved lattice-based designs. 

The use of interleaved lattices can be justified from three perspectives. 
Empirically, we observe that many best-known maximin distance designs in two and three dimensions have interleaved layers. 
Theoretically, the lattice that leads to designs with optimal separation distance as $n \to \infty$, called densest packing, is known for $2\leq p\leq 8$. 
All of them are layered lattices and the densest packings for $p=2,3,4,5,7$ are interleaved lattices~\citep{Conway:1998}. 

Intuitively, consider an arbitrary design $D \subset [0,1]^p$ with size $m$ and 
separation distance $\rho$. 
Clearly, the balls with radius $\rho/2$ that are centered at the design points are non-overlapping and contained in $[-\rho/2,1+\rho/2]^p$. 
Let $\Omega_p$ denote the volume of a unit ball in $\mathbb{R}^p$ and $\delta \leq 1$ denote the volume of the union of these balls divided by the volume of $[-\rho/2,1+\rho/2]^p$, 
then $m \Omega_p (\rho/2)^p = \delta (1+\rho)^p$.
A little derivation yields that 
\[ \rho = 2/\{ (m\Omega_p/\delta)^{1/p} -2 \}, \]
so $\rho$ is high 
if and only if $\delta$ is close to one and $m$ is close to $n$. 
To ensure high $\delta$, the balls around each boundary facet of $[-\rho/2,1+\rho/2]^p$ should be pushed as close to the facet as possible. 
This demands a layer of points be placed on each boundary facet of $[0,1]^p$, 
so layered lattices are desired. 
To further reduce the gap between balls, it is ideal to pack the balls of the second layer in between of the first layer balls,  
so the second layer is desired to be a translation of the first layer. 
Furthermore, it is advantageous to place the third layer balls at the same positions of the first layer balls so that they in turn fill the gaps of the second layer balls. 
Clearly, this calls for interleaved lattices. 

To simultaneously control $\delta$ and $m$, we need to try a variety of interleaved lattices. 
This is because an interleaved lattice leads to designs with high $\delta$ only if it is properly scaled, 
and there may not exist proper $s$ that simultaneously optimizes the scale and $m$. 
For instance, consider $p=2$, $w_1=w_2$ and $L_2$, the lattice generated by \[ G_2 = \left(\begin{array}{cc}
1 & 1 \\ 
0 & 2 
\end{array}\right). \]
Then $b \otimes L_2$ is the two-dimensional densest packing if and only if $b_1/b_2=3^{1/2}$ or $3^{-1/2}$. 
Thus, $\delta$ of $L_2$-based designs is high only if $(s_2-1)/(s_1-1)$ in (\ref{eqn:D}) is close to $3^{1/2}$ or $3^{-1/2}$. 
On the other hand, $m(L_2,s) = \lceil s_1s_2/2 \rceil$. 
As a result, proper $s$ that simultaneously optimizes $(s_2-1)/(s_1-1)$ and $m(L_2,s)$ does not exist for some $n$. 
When proper $s$ does not exist, $L_2$-based designs are poor. 
Fortunately for us, many different types of interleaved lattices that are not necessarily the densest packing lead to designs with high separation distance when they are properly scaled. 
In addition, their optimal scale tends to be diverse, and, as we have discussed, the relationship between $m(L,s)$ and $s$ are quite different for lattices with different $q$ and $r$. 
Consequently, in most cases one or more excellent $(L,s)$ combinations that simultaneously control $\delta$ and $m(L,s)$ exist, 
although we often do not know which is the best before we try many of them. 

We remark that in general interleaved lattice-based designs are only near-optimal. 
In fact, there are many known $(p,n)$ combinations for which interleaved lattice-based designs cannot be optimal in separation distance.

\section{Constructions}
\label{sec:construction}

\subsection{Algorithm 1}
\label{sec:construction1}

In this section, we propose three algorithms to construct interleaved lattice-based maximin distance designs in $[0,1]^p$ with at least $n$ points, where $p\geq 2$ and $n\geq 2$ are given. 
From our first algorithm, we search through all standard interleaved lattices $L$ and all practical span vectors $s$ to find the design in (\ref{eqn:D}) with highest separation distance. 
From Theorem~\ref{thm:sep}, this will produce the optimal interleaved lattice-based design. 
For each lattice, we start with the smallest possible $s$ and gradually increase it. 
In light of Theorems~\ref{thm:m} and~\ref{thm:sep}, we only consider $s$ with $s_k\geq 2$, $k=1,\ldots,p$, and $2^{q-p} \prod_{e_k \notin L} ( 2 \lceil s_k/2 \rceil )$ $\prod_{e_k \in L} s_k \geq n$. 
Let $L_B$, $s_B$ and $\rho_B = \rho(L_B,s_B)$ denote the lattice, span vector and separation distance of the tentatively best solution, respectively.
In light of Theorem~\ref{thm:sep}, 
for each $L$ 
we stop increasing $s$ if $m(L,s)\geq n$ already, 
there is a $k$ such that $e_k\in L$ and $w_k/(s_k-1) \leq \rho_B$, 
there is a $k$ such that $2w_k/(s_k-1) \leq \rho_B$, 
or there is an $x\in L_0$, $x\neq 0_p$ such that $d\{ (s-1_p)^{-1} \otimes x\} \leq \rho_B$. 
Algorithm~1 has six steps below: 

\begin{enumerate}
\item
Obtain the full list of standard interleaved lattices in $p$ dimensions. 
Initialize $\rho_B=0$ and try every lattice $L$.
\item
\label{step:alg1:news} 
For each $L$, initialize $s_1=\cdots=s_{p-1}=2$ and 
\begin{equation}\label{eqn:alg1:sp}
s_p = \max\left[ 2 \left\lceil 2^{p-q-1} n \left\{ \prod_{e_k \notin L, 1\leq k<p} \left( 2 \lceil s_k/2 \rceil \right) \prod_{e_k \in L, 1\leq k<p} s_k \right\}^{-1} \right\rceil -1, 2 \right] .
\end{equation}
\item
\label{step:alg1:afters} 
Compute $m(L,s)$. 
If $m(L,s)<n$, find the smallest $z$ such that $m\{L,(s_1,\ldots,s_{p-1},z)\}\geq n$ and set $s_p=z$. 
\item
\label{step:alg1:ends} 
Compute $\rho(L,s)$ in (\ref{eqn:rho}). If $\rho(L,s) > \rho_B$, update $L_B = L$, $s_B = s$ and $\rho_B = \rho(L_B,s_B)$.  
\item
Find the largest integer $j \leq p$ such that $s_j>2$ and the largest integer $k \leq j-1$ such that $w_k/s_k > \rho_B$ or both $2w_k/s_k > \rho_B$ and $e_k \notin L$.
If no such $j>1$ exists or no such $k$ exists, end the search for the current lattice. 
Otherwise, set $s_k=s_k+1$, $s_l = 2$ for $l=k+1,\ldots,p-1$ and $s_p$ via (\ref{eqn:alg1:sp}) 
and go to Step~\ref{step:alg1:afters}. 
\item
After trying all lattices, output $D(L_B,s_B)$, the best design. 
\end{enumerate}

Algorithm~1 is fast for $p\leq 5$. 
However, because the number of distinct standard interleaved lattices increases dramatically as $p$ grows, it is computationally prohibitive to try every lattice for $p>5$. 
Hence, we recommend to use Algorithm~1 for $2\leq p\leq 5$.

\subsection{Algorithm 2}
\label{sec:construction2}

In this subsection, we propose our second algorithm which is faster than Algorithm~1 for $p>5$. 
To reduce computation, we do not try all standard interleaved lattices. 
Instead, we first find some promising lattices and then focus on designs based on them. 
As discussed in Section~\ref{sec:lattice}, $m(L,s)$ depends on $s$, $q= \log_2 \vert L \cap \{0,1\}^p \vert$ and $r= \vert \{k:e_k \in L\} \vert$ and is almost irrelevant to the specific $L$. 
In light of this, we propose to search through all practical $(s,q,r)$ combinations, each time focusing on one ``best'' lattice.

Clearly, the ``best'' lattice should yield designs with optimal separation distance and near optimal size. 
Such lattices are detected as follows: 
Firstly, let $x^{(k)}$ denote the vector $x$ in $\{e_1,\ldots,e_p\}$ with the $k$th highest $d\{(s-1_p)^{-1}\otimes x\}$. 
To maximize $\min_{e_k \in L}\{w_k/(s_k-1)\}$ in (\ref{eqn:rho}), 
the ``best'' lattice $L$ must contain $x^{(1)},\ldots,x^{(r)}$ but not $x^{(r+1)},\ldots,x^{(p)}$. 
Secondly, sort vectors $x$ in $\Gamma = \{x: x_k=0 \mbox{ or } ( x_k=1 \mbox{ and } e_k\notin L ), k=1,\ldots, p, \sum_{k=1}^p x_k \geq 2\}$ by $d\{(s-1_p)^{-1}\otimes x\}$ 
and decide which of them should be contained in $L_0$ in (\ref{eqn:L0}). 
To maximize $\min_{x\in L_0, x\neq 0_p}d\{(s-1_p)^{-1} \otimes x\}$ in (\ref{eqn:rho}), 
starting from the $x \in \Gamma$ with lowest $d\{(s-1_p)^{-1}\otimes x\}$, we put a vector in $L_0$ unless no proper $L$ with the given $q$ exists. 
We obtain $L_0$ and therefore $L$ after going over all vectors one-by-one. 
From Theorem~\ref{thm:sep}, 
\[ \rho(D) = \min \left[ d\{(s-1_p)^{-1} \otimes x^{(r)}\},d\{(s-1_p)^{-1}\otimes y\},\min_{s_k> 2} \{2w_k/(s_k-1)\} \right], \] 
where $y$ is the first vector in $\Gamma$ that must be put in $L_0$. 
Therefore, we only consider $s$ and $r$ small enough so that $2w_k/(s_k-1)>\rho_B$, $k=1,\ldots,p$, and $d\{(s-1_p)^{-1} \otimes x^{(r)}\} > \rho_B$.
While being optimal in separation distance, the ``best'' lattices tend to attain high second-lowest pairwise distance as well. 

Algorithm~2 has six steps below: 
\begin{enumerate}
\item
Initialize $\rho_B=0$. Try every $q=p,\ldots,1$ from largest to smallest. 
\item
\label{step:alg2:news}
For each $q$, initialize $s_1=\cdots=s_{p-1}=2$ and 
\begin{equation}\label{eqn:alg2:sp}
s_p = \max\left[ 2 \left\lceil 2^{p-q-1} n \left\{ \prod_{k=1}^{p-1} \left( 2 \lceil s_k/2 \rceil \right)\right\}^{-1} \right\rceil -1, 2 \right]. 
\end{equation}
\item
\label{step:alg2:afters} 
Check if there exists $L$ such that $q(L)=q$ and $\{x\in \{0,1\}^p: x\neq 0_p, d\{(s-1_p)^{-1}\otimes x\}\leq \rho_B\} \cap L = \emptyset$. If not, go to Step~\ref{step:alg2:newsp-1}.
\item
For all possible $r$ from largest to smallest, obtain the ``best'' lattice $L$ and compute $m(L,s)$. If $m(L,s)<n$, set $s_p = s_p+1$, break the loop on $r$ and go to Step~\ref{step:alg2:afters}.
Otherwise, update $L_B = L$, $s_B = s$ and $\rho_B = \rho(L_B,s_B)$. 
\item
\label{step:alg2:newsp-1} 
Find the largest integer $j \leq p$ such that $s_j>2$ and the largest integer $k \leq j-1$ such that $2w_k/s_k > \rho_B$. 
If no such $j>1$ exists or no such $k$ exists, end the search for the $q$. 
Otherwise, set $s_k=s_k+1$, $s_l = 2$ for $l=k+1,\ldots,p-1$ and $s_p$ via (\ref{eqn:alg2:sp}) 
and go to Step~\ref{step:alg1:afters}. 
\item
After trying all $q$, output $D(L_B,s_B)$.
\end{enumerate}

Unlike Algorithm~1, from Algorithm~2 we are not guaranteed to find the best interleaved lattice-based design. 
This is because in some rare cases the ``best'' lattice obtained does not have the largest size for the $(s,q,r)$. 
Notwithstanding this fact, Algorithm~2 finds the best interleaved lattice-based design for all scenarios with $2\leq p\leq 5$, $2\leq n\leq 1000$ and equal weight. 
Although Algorithm~2 is fast for $p\leq 8$, it becomes much slower for higher $p$. 
Hence, we recommend to use Algorithm~2 for $6\leq p\leq 8$.

\subsection{Algorithm 3}
\label{sec:construction3}

In this subsection, we propose our third algorithm which is faster than Algorithm~2 for $p>8$. 
In this algorithm, we first generate the design for the eight most important variables using Algorithm~2 and then supplement the remaining dimensions one-by-one, each time greedily finding the design with highest separation distance. 
From our experience, unless $n$ is extremely high and equal weight is used, the best design tends to have only two distinct values, zero and one, for the ninth to the least important variables. 
Hence, we supplement the design using zeros and ones. 
Algorithm~3 has four steps below: 

\begin{enumerate}
\item
Permute the dimensions so that $w_1\geq \cdots \geq w_p$. 
\item
Generate the design $D(L^{(8)},s^{(8)})$ in eight dimensions with at least $n$ points using Algorithm 2. 
\item
For $j$ from 8 to $p-1$, partition $L^{(j)}$ into two sublattices, $L_1$ and $L_2$, where $L_1$ is a lattice, $\mathbb{E}^j \subset L_1 \subset \mathbb{Z}^j$, and $L_2$ is a translation of $L_1$, so that $\rho(L_1,s^{(j)})$ is maximized. 
Let $s^{(j+1)} = (s^{(j)},2)$ and 
$ L^{(j+1)} = \{ (x,2z): x\in L_1, z\in \mathbb{Z} \} \cup \{ (x,2z+1): x\in L_2, z\in \mathbb{Z} \}$.
\item
Obtain $L$ and $s$ by permuting the dimensions of $L^{(p)}$ and $s^{(p)}$, respectively, back to the original order. 
The final design is $D(L,s)$.
\end{enumerate}

Since $L_1$ is determined by $L_1 \cap \{0,1\}^j$, 
in Step~3 we find $L_1$ by assigning the vectors in $( L^{(j)} \cap \{0,1\}^j ) \setminus \{0_j\}$ to either $L_1$ or $L_2$ one-by-one. 
We sort the vectors $x$ from lowest $d\{(s^{(j)}-1_p)^{-1}\otimes x\}$ to highest. 
To maximize $\rho(L_1,s^{(j)})$, 
we attempt to put each $x$ in $L_2$; a vector is put into $L_1$ only if it is necessary. 
Note that if $x$ and $y$ are two vectors in $L_2$, then $x+y$ must be in $L_1$, because otherwise $0_j$ must belong to $L_2$. 
From Theorem~\ref{thm:sep}, $L^{(j+1)}$ obtained in this way has optimal separation distance among designs supplemented from $L^{(j)}$ using 0 and 1. 
We recommend to use Algorithm~3 for $p\geq 9$. 

These algorithms are fast for large $n$. 
For example, from our code that is written purely in R, it takes roughly 2$\cdot$4 minutes to generate the design for $p=20$, $n=1000$ and $w_k=(3/4)^{k-1}$, $k=1,\ldots,p$, using one core of a 2$\cdot$7GHz CPU. 

Both the three algorithms and the algorithm to generate interleaved lattice-based minimax distance designs~\citep{ILmMD} search through a variety of lattices and span vectors to find a best design. 
In both problems, we cannot afford to try all standard interleaved lattices unless $p$ is very small. 
Different strategies are adopted to reduce the number of lattices in the search space. 
In \citet{ILmMD}, dimension permuted lattices are treated as the same lattice, which is only suitable under the equal weight assumption. 
Even so, the algorithm becomes slow for $p>8$. 
In our Algorithm~2, we find one best lattice for given $(s,q,r)$. 
This strategy cannot be applied to the minimax distance design problem, either, 
because there is no simple way to tell which lattice yields designs with lowest fill distance. 
Consequently, although producing designs with similar structures, algorithms in this paper are quite different from that in \citet{ILmMD}.

\section{Numerical comparison}\label{sec:simu}

In this section, we compare interleaved lattice-based maximin distance designs to best-known maximin distance designs for $p=2$ and $3$, 
maximin distance designs generated from \citet{Stinstra:2003} using its SFDP** formulation, and maximin distance Latin hypercube designs. 
We tune the latter two methods to unequal weights: 
We first generate optimal designs in $\prod_{k=1}^p [0,w_k]$ from \citet{Stinstra:2003} and then scale them to $[0,1]^p$; 
we use the weighted distance measure (\ref{eqn:distance}) to generate maximin distance Latin hypercube designs. 

\begin{figure} 
\begin{center}
\includegraphics[width=7.32cm]{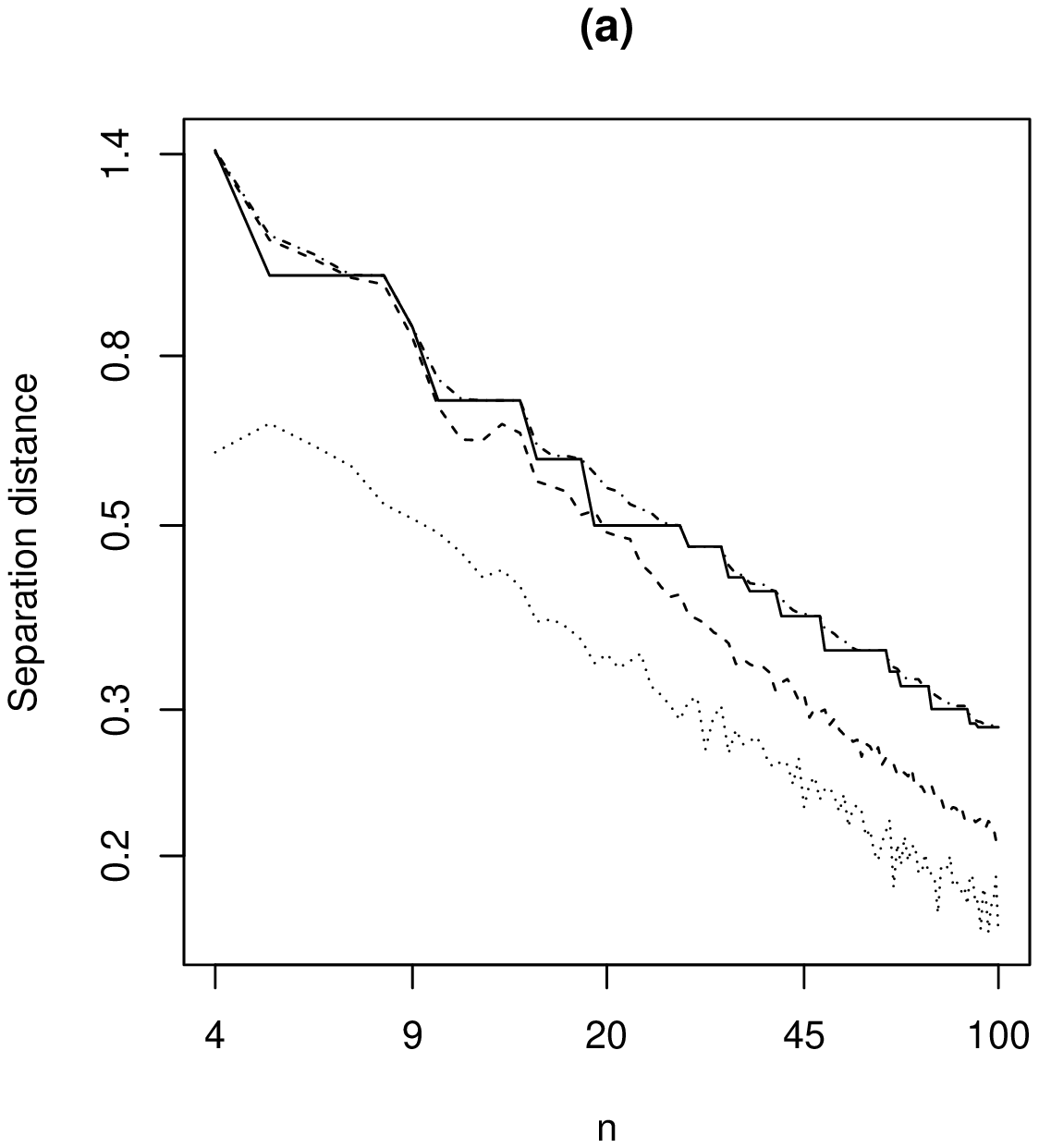}
\includegraphics[width=7.32cm]{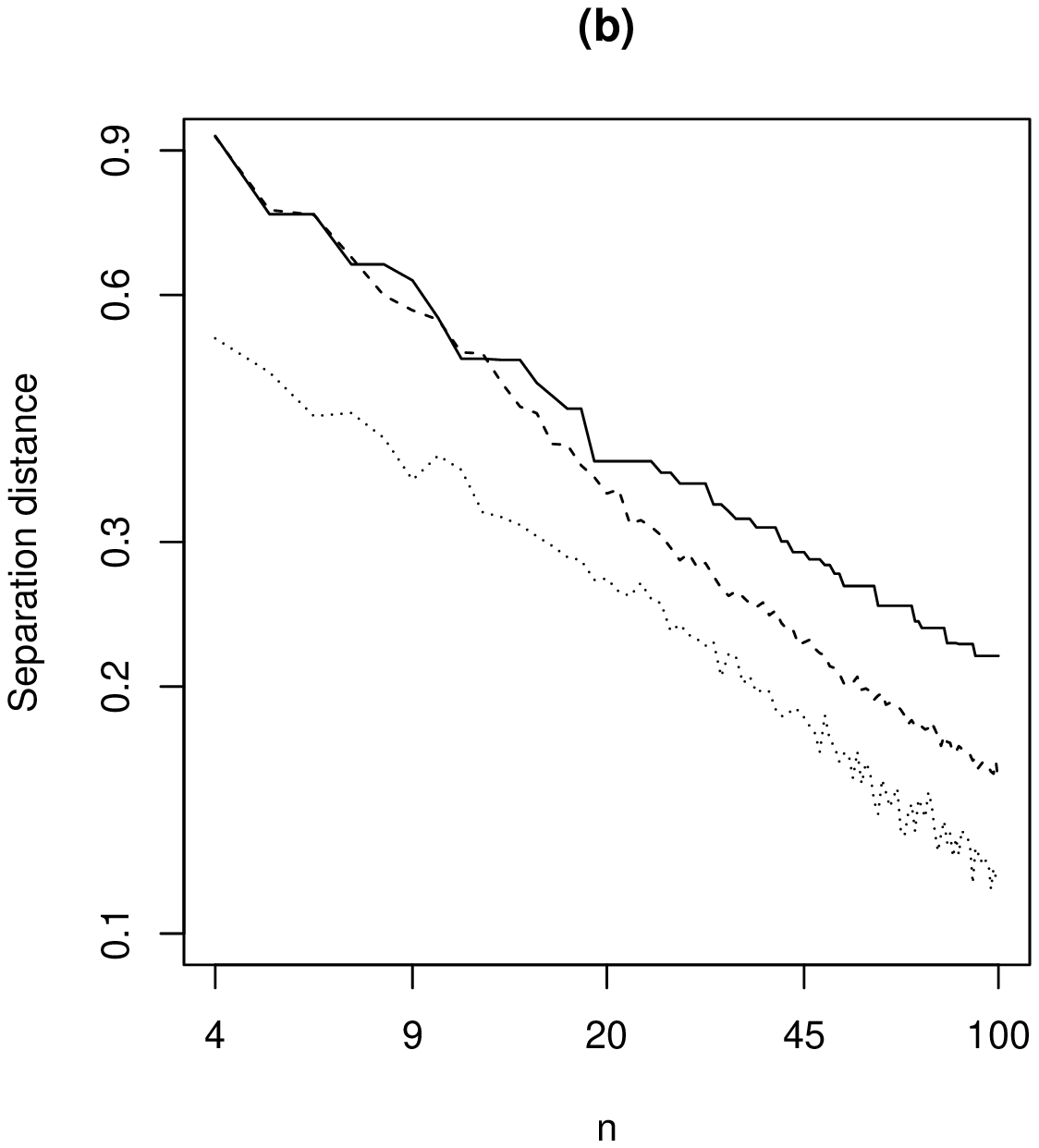}

\includegraphics[width=7.32cm]{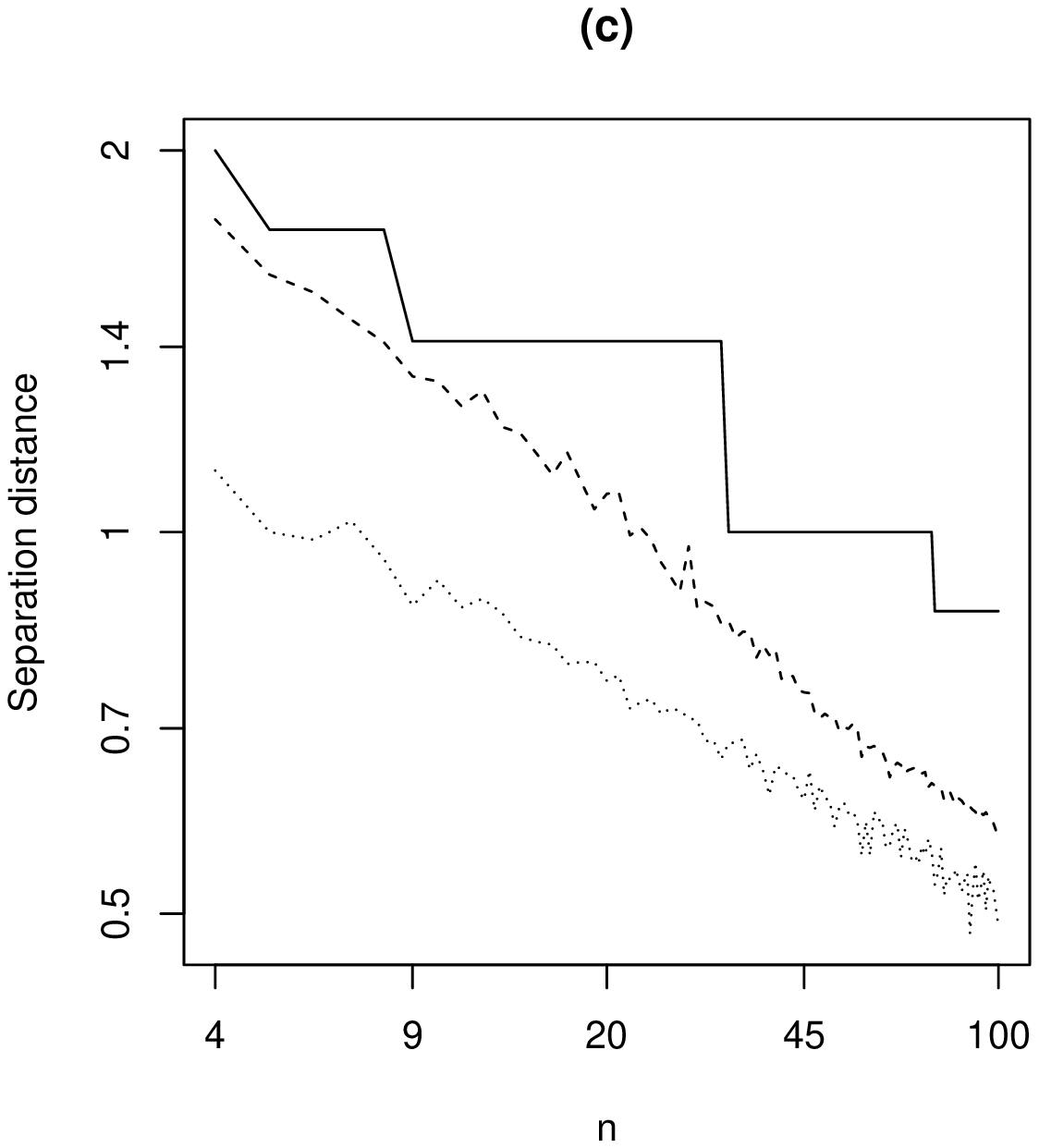}
\includegraphics[width=7.32cm]{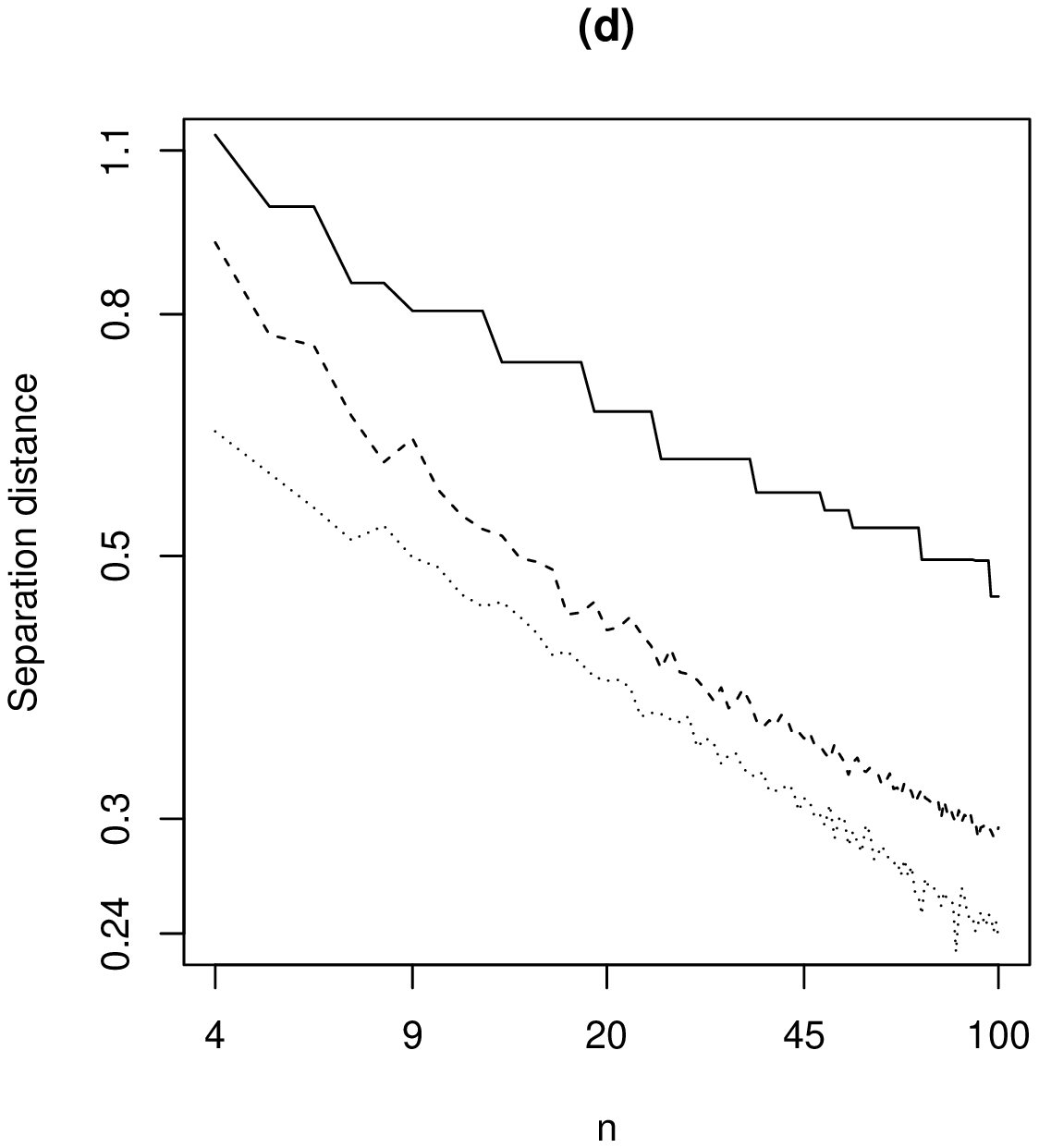}
\caption{Separation distances of four types of designs, interleaved lattice-based maximin distance designs (solid), maximin distance designs by~\citet{Stinstra:2003} (dashed), maximin distance Latin hypercube designs (dotted), and best-known maximin distance designs (dash-dotted) in (a) three dimensions with equal weights, (b) three dimensions with unequal weights, (c) six dimensions with equal weights, and (d) fifteen dimensions with unequal weights.}\label{fig:sep}
\end{center}
\end{figure}

We first compare the separation distance of these designs. 
Figure~\ref{fig:sep} presents the results in four scenarios: (a) $p=3$ with $w_1=w_2=w_3=1$, (b) $p=3$ with unequal weights $w_k=(3/4)^{k-1}$, $k=1,\ldots,p$, (c) $p=6$ with $w_1=\cdots=w_6=1$, and (d) $p=15$ with unequal weights $w_k=(3/4)^{k-1}$, $k=1,\ldots,p$. 
We remark that for some $(p,n)$ combinations our method generates a design with $m> n$ points.
Because we can obtain an $n$-point design with the same separation distance by simply removing $m-n$ points, the separation distance for $(p,n)$ is the same as that for $(p,m)$.  
For $p=3$ with equal weights, our proposed designs are only slightly inferior to best-known designs, indicating that they are near-optimal for low $p$. 
In all scenarios, our proposed designs have much higher separation distance than maximin distance designs generated from \citet{Stinstra:2003} and maximin distance Latin hypercube designs. 
This suggests that for $p\geq 4$, unconstrained numerical search or numerical search constrained on Latin hypercubes cannot produce near-optimal solutions, and our proposed designs are by far the best maximin distance designs. 
From our experience, our proposed designs are usually better than other designs by at least 0$\cdot$1 for $p\geq 4$. 

Next, we compare our proposed designs with maximin distance Latin hypercube designs on integrated mean squared prediction error, 
$ \int_{[0,1]^p} E [ \{ \hat Y(x) - Y(x) \}^2 ] dx $, 
where $Y(x)$ is the realization of a Gaussian process and $\hat Y(x)$ is the predicted outcome from a Gaussian process model with correctly specified covariance function. 
We assume the Gaussian process has constant mean and the covariance between $x$ and $y$ is 
$ \exp \{ -\theta \sum_{k=1}^p (v_k|x_k-y_k|)^2 \} $,
where $\theta=10$ and $v=(v_1,\ldots,v_p)$ are known after experimentation. 
Because the integrated mean squared prediction error criterion is sensitive to whether the design has disproportionally denser points around the boundary, for a fair comparison we use transformed designs in (\ref{eqn:tildeD}) for our method. 

\begin{figure} 
\begin{center}
\includegraphics[width=7.32cm]{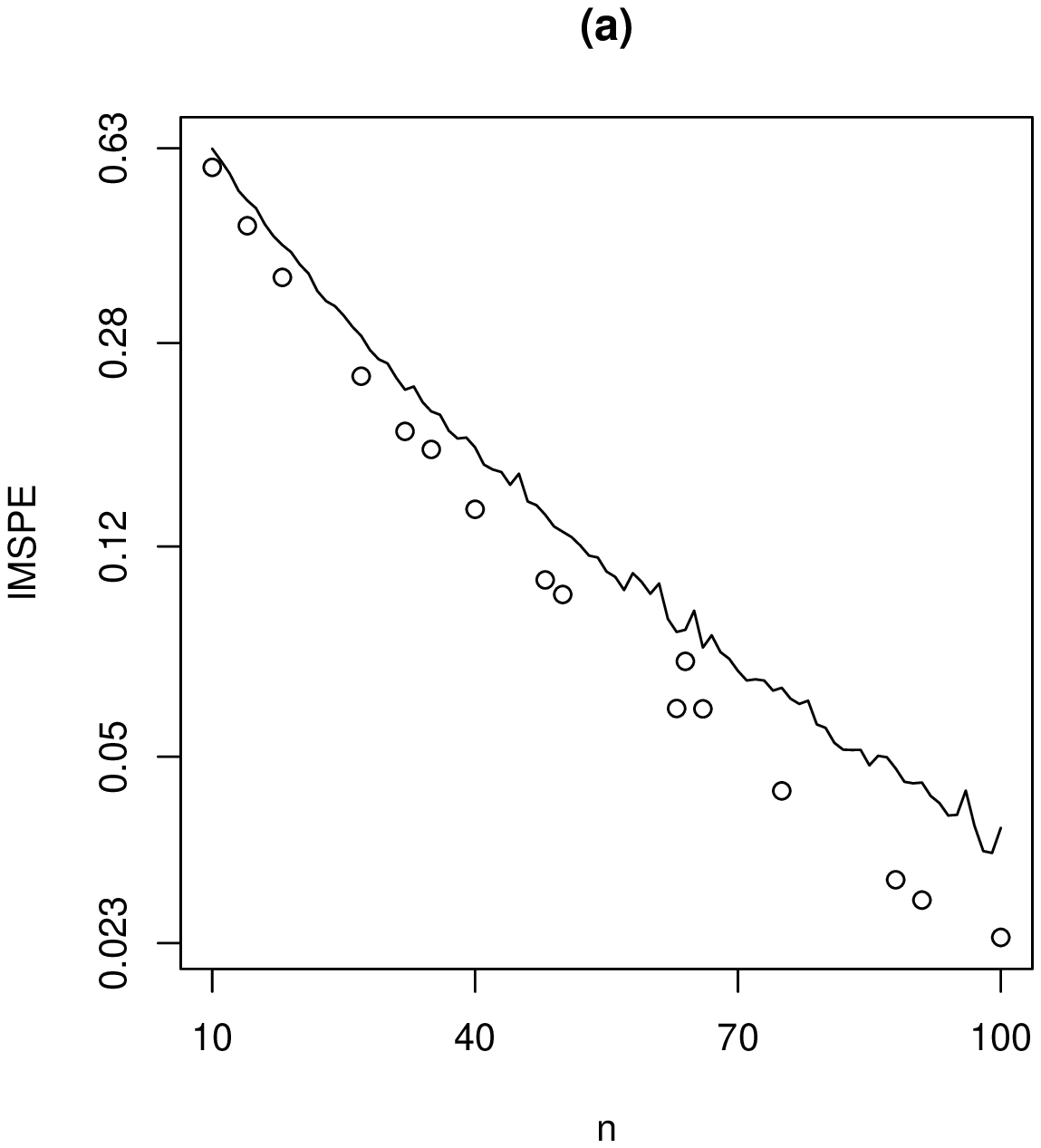}
\includegraphics[width=7.32cm]{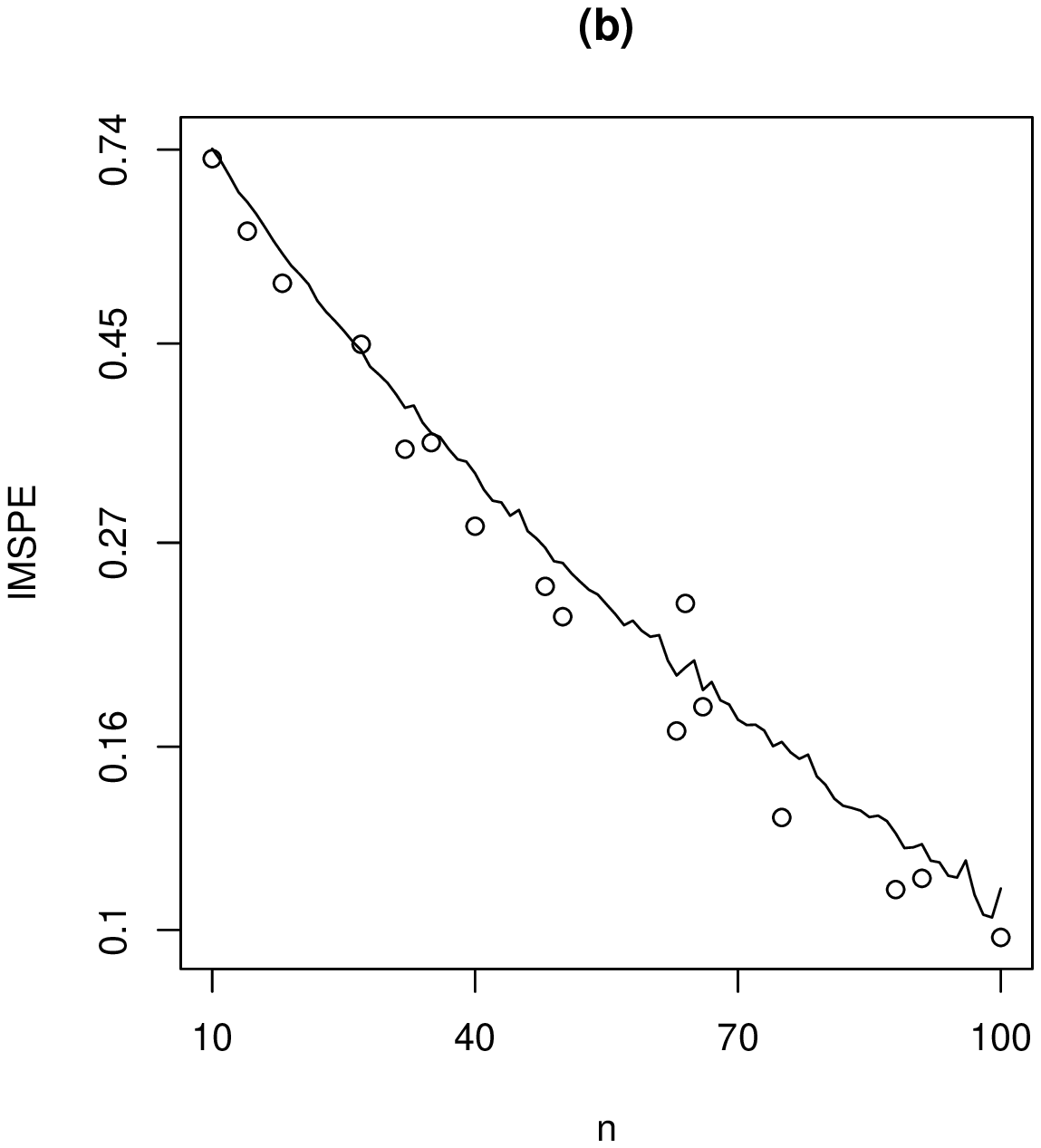}

\includegraphics[width=7.32cm]{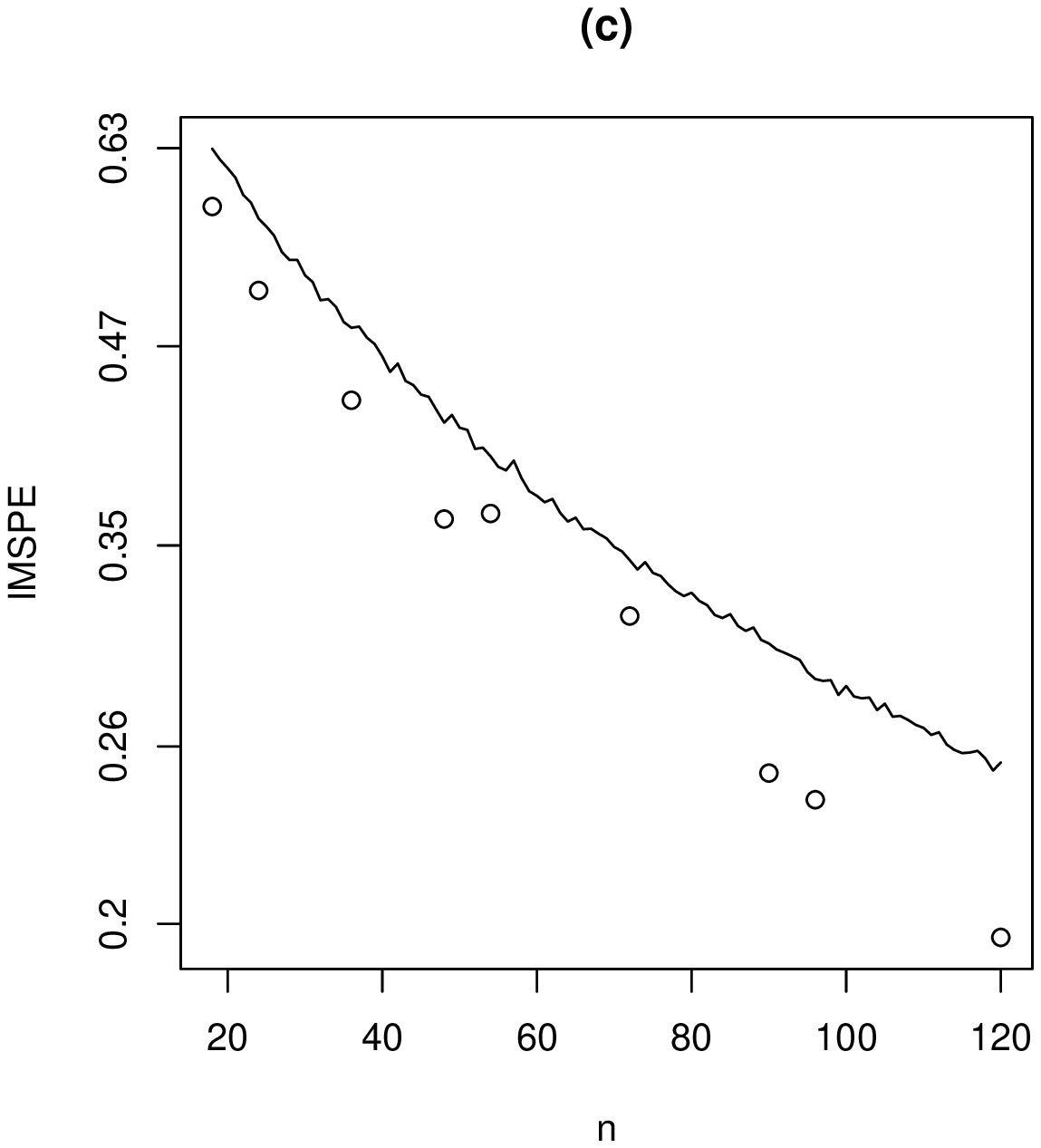}
\includegraphics[width=7.32cm]{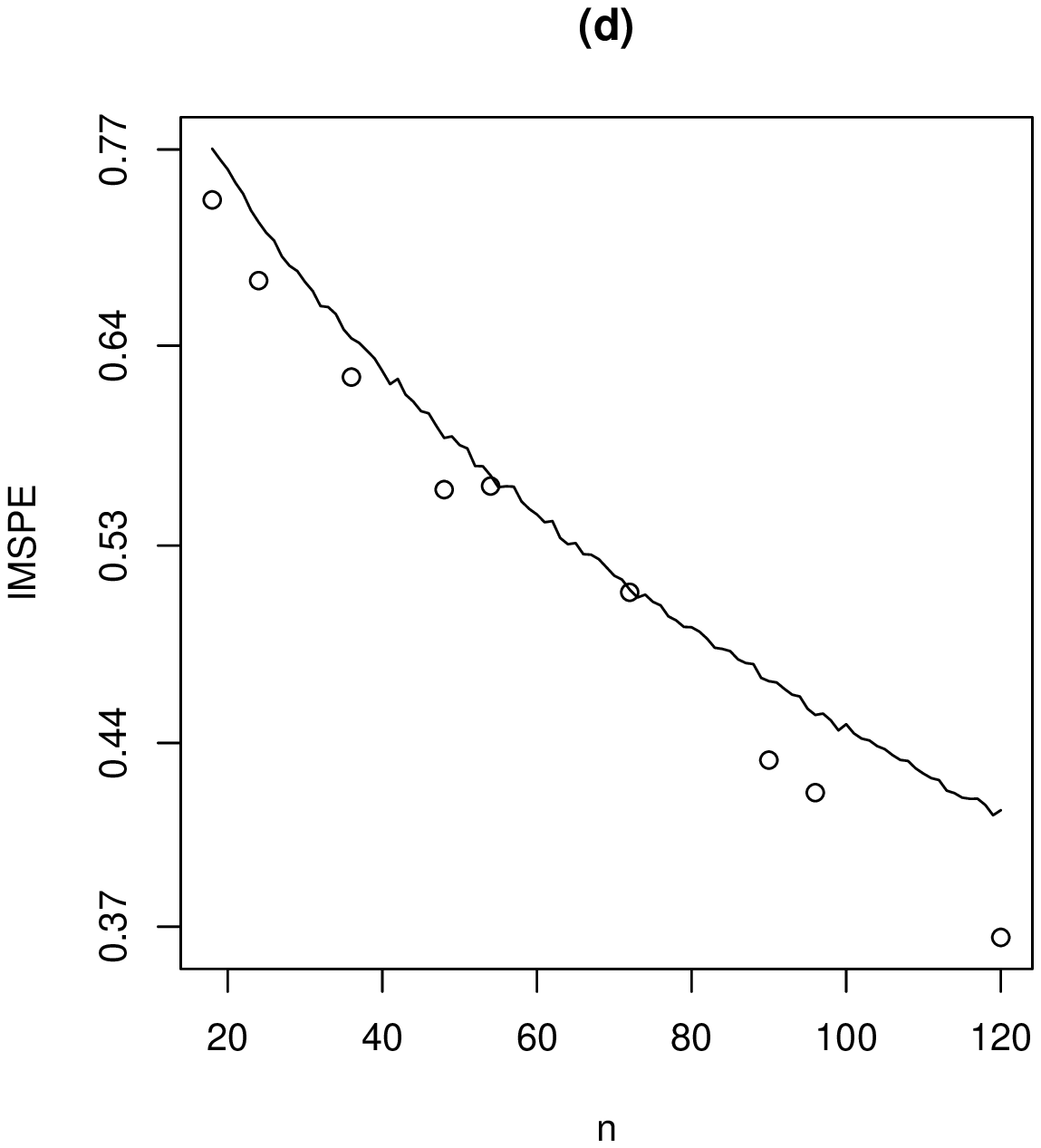}
\caption{Integrated mean squared prediction error of two types of designs, interleaved lattice-based maximin distance designs (circles) and maximin distance Latin hypercube designs (solid line) in (a) three dimensions with correct equal weights, (b) three dimensions with roughly correct equal weights, (c) eight dimensions with correct unequal weights, and (d) eight dimensions with roughly correct unequal weights.} \label{fig:IMSPE}
\end{center}
\end{figure}

We consider four scenarios.  
In the first and second scenarios, we assume $p=3$ and designs are generated using $w_1=w_2=w_3=1$. 
In the third and fourth scenarios, we assume $p=8$ and designs are generated using $w_k=(3/4)^{k-1}$, $k=1,\ldots,8$. 
In the first and third scenarios, we assume that $v=w$, i.e., 
we use correct weights to generate designs. 
In the remaining two scenarios, we assume that $v_k=w_k u_k$, $k=1,\ldots,p$, where the $u_k$'s are independently sampled from the uniform distribution on $[1/2,2]$, i.e., 
we use roughly correct weights to generate designs. 
In these two scenarios, we independently generate 20 sets of $v$ and record the averaged results. 
These assumptions represent the situations in which the variable importance is exactly or roughly known before experimentation. 
Figure~\ref{fig:IMSPE} presents numerical comparison results on integrated mean squared prediction error. 
For our method we only include $(p,n)$ combinations for which the generated design has exactly $n$ points. 
Our proposed designs outperform maximin distance Latin hypercube designs in all scenarios, 
showing that our method is useful when we have precise or relatively accurate prior knowledge on the variable importance. 
The gap in performance is less favorable to our proposed designs for roughly correct weights than for exactly correct weights, 
implying that our proposed designs are more competitive when more accurate information on variable importance is available. 

While possessing excellent separation distance, from our experience our proposed designs also have reasonably good fill distance. 
Besides, from our proposed designs each variable has at least two levels, 
and more important variables are usually assigned with more levels, 
allowing estimation of linear main effects for unimportant variables and estimation of linear and higher order main effects for important variables. 

One future problem is to extend our method to computer experiments with mixed continuous, ordinal and categorical variables. 
It is also interesting to construct designs that simultaneously possess high separation distance and low fill distance.

\appendix
\section{Proofs}
\label{sec:proof}

\begin{proof}[Proof of Theorem~1]
Clearly, there exist $u \in \mathbb{Z}^p$ and $\bar c \in \mathbb{R}^p$ such that 
\[ D = b \otimes \left\{ L \cap \left( \prod_{k=1}^p [u_k, u_k+s_k-1] \right) \right\} \oplus \bar c. \]
Because $L$ is a lattice and $\mathbb{E}^p \subset L \subset \mathbb{Z}^p$, for any finite set $A\subset \mathbb{Z}^p$, we have
\[ m( L \cap A ) = m\{ L \cap ( A \oplus e_k ) \} \] 
for any $k$ such that $e_k \in L$ and 
\[ m( L \cap A ) = m[ L \cap \{ A \oplus (2e_k) \} ] \] 
for $k=1,\ldots,p$. 
Therefore, 
$ m(L_0) 2^{r(L)} = 2^{q(L)}$, and 
\[ m(D) = m\left\{ L \cap \left( \prod_{k=1}^p [ u_k, u_k + z_k -1 ] \right) \right\}
 = m(L_0) \prod_{e_k \notin L} (z_k/2) \prod_{e_k\in L} z_k  \]
for any $z \in \mathbb{N}^p$ and $z_k \in \mathbb{E}$ for any $k$ with $e_k \notin L$. 
Therefore, 
\[ 2^{q-p} \prod_{e_k \notin L} \left( 2 \lfloor s_k/2 \rfloor \right) \prod_{e_k\in L} s_k \leq m(D) \leq 2^{q-p} \prod_{e_k \notin L} \left( 2 \lceil s_k/2 \rceil \right) \prod_{e_k \in L} s_k . \]

For any $K \subset \{1,\ldots,p\}$, let $s_K=(s_{K,1},\ldots,s_{K,p})$ be the $p$-vector with $s_{K,k} = 2$ for $k\in K$ and $s_{K,k} = 1$ for $k\notin K$.  
Then 
\[ m(D) = m(L,s,u) = \sum_{K \subset \{1,\ldots,p\}} \left\{ m(L, s_K, u) \prod_{k\in K} \lfloor s_k/2 \rfloor \prod_{k \notin K} (s_k - 2 \lfloor s_k/2 \rfloor) \right\}. \]

For any $p$-vector $u$, let $u_K$ denote the projection of $u$ unto the dimensions in $K$. 
Let 
\[ L_K = \{ u_K : u \in L, u_k = 0 \mbox{ for any } k \notin K \}. \]
Clearly, for any $K$, $L_K$ is a lattice. If there exists a $v\in L$ such that $v_k=u_k$ for any $k \in K$, $m(L, s_K, u) = q(L_K)$; otherwise $m(L, s_K, u)=0$. 
Therefore, $m(L, s_K, 0_p) = q(L_K) = \sup_u m(L, s_K, u)$. 
Therefore, 
\[ m(L,s,0_p) = \sup_u m(L,s,u), \]
which completes the proof. 
\end{proof}

\begin{proof}[Proof of Theorem~2]
Consider an arbitrary design $\bar D$ which is generated from $\bar L$, $\bar b$ and $\bar c$ with $m(\bar D)\geq n$. 
We shall show that based on $\bar D$ we can always find its counterpart, $\grave D$, which is generated from $\grave L$, $\grave b$ and $\grave c$ such that $\grave L$ is a standard interleaved lattice, $\grave s_k\geq 2$, $\grave b_k=1/(\grave s_k-1)$ and $\grave c_k=0$ for $k=1,\ldots,p$, $\rho(\grave D)\geq \rho(\bar D)$, and $m(\grave D)\geq m(\bar D)$.
 
Let $\bar s$ denote the span vector of $\bar D$. 
Without loss of generality, assume $\bar s_k \geq 2$ for $1\leq k\leq h$, $\bar s_k = 1$ for $h+1 \leq k \leq p$, and $h\in \{0,\ldots,p\}$. Because $m(\bar D)\geq 2$, $h\geq 1$. 
Let $l_k$ denote the lowest $k$-dimensional value of $\bar D$, $k=1,\ldots,p$. 
When $h<p$, let $\bar a_k = (l_k-\bar c_k)/\bar b_k$, $h+1\leq k\leq p$ and 
\[ \bar L^* = \{ z\in \mathbb{Z}^h: (z,\bar a_{h+1},\ldots,\bar a_p) \in \bar L \}. \]
Let $y \in \{0,1\}^h$ be an arbitrary element of $\bar L^*$ and $\bar L^{**} = \bar L^* \oplus y$. 
When $h=p$, let $\bar L^{**} = \bar L$. 
Then $\bar L^{**}$ is a lattice and $\mathbb{E}^h \subset \bar L^{**} \subset \mathbb{Z}^h$. 
%Let $H = \{ z\in\mathbb{Z}^p : \sum_{k=1}^p z_k \mbox{ is an even integer} \}$ be a standard interleaved lattice. 

Now define $\acute L$, $\acute b$ and $\acute c$ as follows: 
If $L^{**} = \mathbb{E}^h$, let $\acute L = \mathbb{Z}^p$ and $\acute b_k = 2\bar b_k$, $k=1,\ldots,h$.  
If $L^{**} \neq \mathbb{E}^h$, let $\acute L = L^{**} \times \mathbb{Z}^{p-h}$ and $\acute b_k = \bar b_k$, $k=1,\ldots,h$.  
Let $\acute c_k = l_k + \acute b_k y_k$, $k=1,\ldots,h$, and $\acute b_k = 2$ and $\acute c_k = l_k$, $k=h+1,\ldots,p$. 
Then $\bar D$ can be seen as generated from $\acute L$, $\acute b$ and $\acute c$, $\acute L$ is a lattice and $\mathbb{E}^p \subset \acute L \subset \mathbb{Z}^p$. 
Let $\acute D$ be the design generated from $\acute L$, $\acute b$ and $l$. 
Clearly, $\rho(\acute D) = \rho(\bar D)$. 
From Theorem~1, $m(\acute D) \geq m(\bar D)$. 

Next, let $\check b$ be the $p$-vector such that $\check b_k = 1/(\bar s_k-1)$, $k=1,\ldots,h$ and $\check b_k = 2$, $k=h+1,\ldots,p$. 
Consider the design $\check D$ that is generated from $\acute L$, $\check b$ and $0_p$. 
Clearly, $m(\check D) = m(\acute D)$ and $\rho(\check D) \geq \rho(\acute D)$. 

Finally, let 
\[ \check L^* = \{ z\in \mathbb{Z}^h: (z,0_{p-h}) \in \check L \}. \]
Clearly, $\check L^*$ is a lattice, $\mathbb{E}^h \subset \check L^* \subset \mathbb{Z}^h$ and $\check L^* \neq \mathbb{E}^h$. 
Partition $\check L^*$ into two sublattices, $\check L_1$ and $\check L_2$, such that $\mathbb{E}^h \subset \check L_1 \subset \mathbb{Z}^h$ and $\check L_2$ is a translation of $\check L_1$. 
To do so, assume $\check L^* \cap \{0,1\}^h$ has $2^{\check q}$ elements. 
Clearly, $\check q\geq 1$. 
Select arbitrary $\check q-1$ elements of $\check L^* \cap \{0,1\}^h$, $v_1,\ldots,v_{\check q-1}$, such that 
$\sum_{i=1}^{\check q-1} ( a_i v_i ) \notin \mathbb{E}^h$ for any $(a_1,\ldots,a_{\check q-1}) \in \{0,1\}^{\check q-1} \setminus \{0_{\check q-1}\}$. 
Letting $\check L_1$ be the space generated from $v_1,\ldots,v_{\check q-1}, 2e_1,\ldots,2e_h$ and $\check L_2 = \check L^* \setminus \check L_1$, the $\check L_1$ and $\check L_2$ satisfy our requirements. 
Let 
\[ \grave L = (\check L_1 \times \mathbb{E}^{p-h} ) \cup \{ \check L_2 \times ( \mathbb{E}^{p-h} \oplus 1_{p-h} ) \}, \]
$\grave b$ be the $p$-vector such that $\grave b_k = \check b_k$, $k=1,\ldots,h$ and $\grave b_k = 1$, $k=h+1,\ldots,p$ and $\grave D$ be the design generated from $\grave L$, $\grave b$ and $0_p$. 
Then $\grave L$ is a standard interleaved lattice, $\rho(\grave D) \geq \rho(\check D)$ and $m(\grave D)=m(\check D)$. 
Let $\grave s_k$ denote the number of distinct values of the $k$-th dimension of $\grave D$. 
Then $\grave s_k\geq 2$ and $\grave b_k=1/(\grave s_k-1)$ for $k=1,\ldots,p$. 

Since $\bar D$ can be any interleaved lattice-based design, there exist at least one optimal interleaved lattice-based design that satisfy the properties listed in the theorem. 

Clearly, for such $D$ 
\[ \rho(D) \leq \min\left\{ \min_{z\in L_0, z\neq 0_p} d(b \otimes z), \min_{e_k\in L} (b_k w_k), \min_{s_k> 2} (2b_k w_k) \right\}. \]
Consider an arbitrary pair of points $y,z\in L$. 
Firstly, if there is a $k$ such that $|z_k-y_k|\geq 2$. 
Then $s_k >2$ and $d(z_k-y_k) \geq 2 b_k w_k > b_k w_k$. 
Secondly, if there is a $k$ such that $|z_k-y_k|\geq 1$ and $e_k\in L$. 
Then $d(z_k-y_k) \geq b_k w_k$. 
Finally, if $|z_k-y_k|= 0$ for any $e_k\in L$ and $|z_k-y_k|\leq 1$ for any $k$. 
Then $d(z_k-y_k) \geq \min_{z\in L_0, z\neq 0_p} d(b \otimes z)$. 
Combining the three cases, \[ d(z_k-y_k) \geq \min\left\{ \min_{z\in L_0, z\neq 0_p} d(b \otimes z), \min_{e_k\in L} (b_k w_k), \min_{s_k> 2} (2 b_k w_k) \right\}, \]
which completes the proof. 
\end{proof}

\begin{proof}[Proof of Theorem~3]
Let $L$ be an arbitrary lattice $L$ with $\mathbb{E}^p \subset L \subset \mathbb{Z}^p$, $q(L)=z_3$ and $r(L)=z_1$. 
We shall find a counterpart of $L$, $\tilde L$, such that $q(\tilde L)=z_3$, $r(\tilde L)=z_2$ and $m(\tilde L,s)\geq m(L,s)$ for any $s\in \mathbb{N}^p$. 

Without loss of generality, assume $e_k \in L$, $k=p-z_2 +1,\ldots,p-z_2+z_1$ and $e_k \notin L$, $k=1,\ldots,p-z_2, p-z_2+z_1+1,\ldots,p$. 
Let 
\[ L^*(h) = \{ v \in \mathbb{Z}^{p-z_2}: (v,h) \in L \}. \]
Clearly, $L^*(0_{z_2})$ is a lattice, $\mathbb{E}^{p-z_2} \subset L^*(0_{z_2}) \subset \mathbb{Z}^{p-z_2}$ and $q\{L^*(0_{z_2})\} \geq q(L)-z_2$. 
If $q\{L^*(0_{z_2})\} > q(L)-z_2$, let $L^{**}$ by an arbitrary sublattice of $L^*(0_{z_2})$ such that $\mathbb{E}^p \subset L^{**} \subset \mathbb{Z}^p$ and $q\{L^{**}\} = q(L)-z_2$. 
Otherwise let $L^{**} = L^*(0_{z_2})$. 
Let $\tilde L = L^{**} \times \mathbb{Z}^{z_2}$, 
then $\tilde L$ is a lattice, $\mathbb{E}^p \subset \tilde L \subset \mathbb{Z}^p$, $q(\tilde L)=q(L)$ and $r(\tilde L) = z_2$. 
From Theorem~1, $L^*(h)$ has maximum number of points when $h=0_{z_2}$. 
Therefore, $m(L,s)\leq m(\tilde L,s)$ for any $s\in \mathbb{N}^p$. 
\end{proof}

\bibliographystyle{Chicago}
\bibliography{ILMaximin}

\end{document}